\documentclass[aps,prl,twocolumn,superscriptaddress,showpacs]{revtex4}

\usepackage{amsmath,amssymb,bm}
\usepackage{graphicx}

\begin{document}

\title{Electrical Switching in Metallic Carbon Nanotubes}
\author{Young-Woo Son}
\affiliation{Department of Physics, University of California at Berkeley 
and Materials Sciences Division at Lawrence Berkeley National Laboratory, CA 94720}
\affiliation{School of Physics, Seoul National University, Seoul
  151-742, Korea}
\author{Jisoon Ihm}
\affiliation{School of Physics, Seoul National University, Seoul
  151-742, Korea}
\affiliation{Department of Physics, Sungkyunkwan University, Suwon 440-746, Korea}
\author{Marvin L. Cohen}
\affiliation{Department of Physics, University of California at Berkeley 
and Materials Sciences Division at Lawrence Berkeley National Laboratory, CA 94720}
\author{Steven G. Louie}
\affiliation{Department of Physics, University of California at Berkeley 
and Materials Sciences Division at Lawrence Berkeley National Laboratory, CA 94720}
\author{Hyoung Joon Choi}
\email[Email:\ ]{h.j.choi@yonsei.ac.kr}
\affiliation{School of Computational Sciences, Korea Institute for
Advanced Study, Seoul 130-722, Korea, and Institute of Physics and Applied Physics, 
Yonsei University, Seoul 120-749, Korea}

\begin{abstract}
We present first-principles calculations of quantum transport which show 
that the resistance of metallic carbon nanotubes can be changed dramatically 
with homogeneous transverse electric fields if the nanotubes have 
impurities or defects. 
The change of the resistance is predicted to range over more than two orders 
of magnitude with experimentally attainable electric fields. 
This novel property has its origin that backscattering of conduction 
electrons by impurities or defects in the nanotubes is strongly dependent 
on the strength and/or direction of the applied electric fields. 
We expect this property to open a path to new device applications 
of metallic carbon nanotubes.
\end{abstract}

\maketitle

Are single-walled carbon nanotubes (SWNTs) suitable 
for future nano-electronic device applications? 
The answer depends on whether the electrical 
properties of carbon nanotubes are changeable by applied 
gate voltages or electric fields~\cite{mceuen}. 
Carbon nanotubes are either semiconducting or metallic, 
depending on their atomic geometry (diameter and chirality). 
Semiconducting carbon nanotubes are well suited for 
nano-electronic device applications~\cite{tans,soh,martel,javey} 
because their electrical resistance is controllable 
by the gate voltage just as in silicon-based field-effect transistors. 
On the other hand, metallic nanotubes reportedly have electrical 
resistances which are not sensitive to the gate voltage or 
homogeneous transverse electric fields~\cite{bockrath,nygard,okeeffe,li} 
and this insensitivity has discouraged device applications 
of the metallic ones. 
The previous reports on metallic nanotubes, however, are limited 
to clean ones with electric fields~\cite{okeeffe,li,novikov,yhkim,barraza} 
or to defective ones without electric fields~\cite{choi2,kostyrko,ando,zhao}
even though impurities or structural defects under the transverse
electric field may produce exotic effects because of the 
low dimensionality of the nanotubes.

A clean armchair-type SWNT is metallic with two linear bands 
intersecting at the Fermi energy ($E_F$) regardless of its diameter~\cite{saito} (Fig. 1(a)). 
The tube's electrical resistance is determined by its electronic structure. 
A clean metallic SWNT should have an electrical resistance of 6.5 k$\Omega$ 
in two-probe measurements with perfect electrical contacts~\cite{saito,frank}. 
This results from the resistance quantum, 12.9 k$\Omega$~\cite{landauer,fisher} 
(which is $h/2e^2$ with $h$ = Planck's constant and $e$ = charge of an electron) 
divided by the number of bands at ${E}_F$. 
The resistance of a clean metallic carbon nanotube is insensitive 
to a homogeneous transverse (i.e., perpendicular to the tubular axis) 
electric field. 
Although the applied electric field polarizes the nanotube along the field 
direction~\cite{benedict} and the band dispersion is modified near ${E}_F$
(Fermi velocities are slightly decreased as shown in Fig. 1(b)), 
electric fields of moderate strength do not change 
the number of bands at ${E}_F$~\cite{okeeffe,li,novikov,yhkim} 
which is the only material parameter that determines the resistance of a 
clean one dimensional sample. 
Contrary to the clean tube case, however, a defective metallic carbon nanotube, 
as will be shown in this Letter, will have an electrical resistance which is very 
sensitive to transverse electric fields, and this property may be exploited 
for switching device applications.

\begin{figure}[b]
\centering
\includegraphics[width=8.5cm]{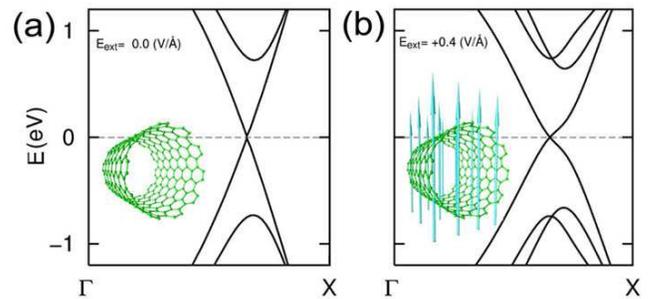}
\caption{Electronic structure of a clean SWNT with and without 
a transverse electric field ($E_\text{ext}$). 
The Fermi level is set to zero in all figures. 
(a) Band structure near the Fermi level of a clean (10, 10) 
SWNT with $E_\text{ext}$ = 0. 
The inset is the atomic structure of the (10, 10) SWNT. 
(b) Band structure of a clean (10, 10) SWNT with 
$E_\text{ext}$ = 0.4 V/\AA. 
The inset shows the nanotube and the applied electric field 
perpendicular to the tubular axis.}
\end{figure}

Our study of the resistance of defective metallic SWNTs in transverse 
electric fields (Fig. 2) is based on first-principles calculations~\cite{soler}. 
We introduce either impurities or structural defects into metallic nanotubes 
and describe their atomic and electronic structures using norm-conserving
pseudopotentials~\cite{troullier} with Kleinman-Bylander's nonlocal
projectors~\cite{kleinman} and the local density approximation for the 
exchange-correlation potential.
The wavefunction is expanded with a single zeta basis set~\cite{soler} to 
produce the similar relaxed geometric configurations with the defect 
and the electronic structures compared with the result
using plane-wave basis set~\cite{choi2}.
A periodic saw-tooth-type potential perpendicular to the direction 
of the tube axis is used to simulate the applied transverse electric field 
in a supercell having 420 carbon atoms for (10,10) nanotube and 576 atoms 
for (18,0) tube respectively
and such a potential is homogeneous along the tube axis 
as shown in the inset of Fig. 1(b).
By calculating the scattering wavefunctions of electrons around the defects, 
we obtain the quantum-mechanical probability for an electron near ${E}_F$
to transmit through the defects~\cite{choi1,yyoon}. 
The two-probe resistance of the sample is inversely proportional to the 
obtained transmission probability as well as the number of bands~\cite{datta}. 
Our calculational results show that the resistance of defective metallic SWNTs 
is very sensitive to the strength and/or direction of the applied transverse 
electric field (Figs. 2-4).

\begin{figure}[t]
\centering
\includegraphics[width=8.0cm]{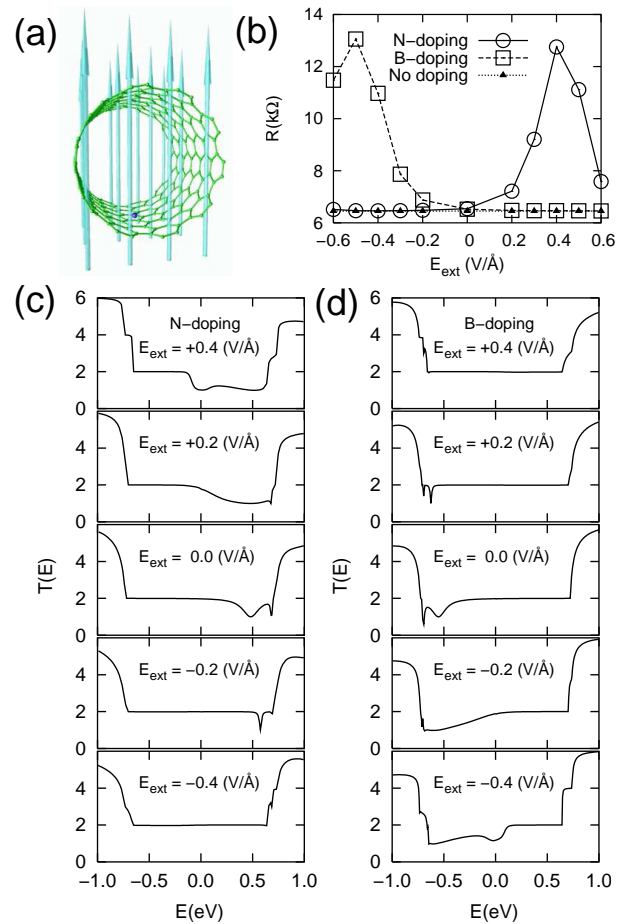}
\caption{Electrical resistances of single-impurity doped SWNTs in 
transverse electric field $E_\text{ext}$. 
(a) A ball-and-stick model of a nitrogen- or boron-doped SWNT 
in a transverse electric field. 
The impurity is denoted by a darker atom. 
The direction of the arrows (from the impurity-doped side to the other) 
corresponds to $E_\text{ext}> 0$. 
(b) Calculated two-probe electrical resistances ($R$) 
of a nitrogen-doped (10,10) SWNT, a boron-doped one and a clean one. 
Transmission spectra  $T(E)$  of (c) a nitrogen-doped 
and (d) a boron-doped (10,10) SWNT in various transverse electric 
fields.}
\end{figure}

Impurities or structural defects in metallic nanotubes produce quasibound 
states which backscatter conducting electrons resonantly near the quasibound-state energies.
In case of a nitrogen and a boron doped (10, 10) nanotube (Fig. 2(a)), the electrical 
resistance shows maximum values at the electric field of 0.4 V/\AA~and 
$-$0.5 V/\AA~, respectively (Fig. 2(b)). 
Without electric fields, the nitrogen (boron) impurity produces resonant 
backscattering dips in the transmission associated with donor-like (acceptor-like) 
quasibound states above (below) ${E}_F$ as shown 
in the middle panels (denoted by $E_\text{ext}=0.0$V/\AA) of Fig. 2(c-d)~\cite{choi2}. 

When transverse electric fields are applied, the field-induced change of 
the screened electrostatic potential near the impurity sites changes 
the quasibound-state energies and thereby the resonant backscattering 
and the resistance of the nanotubes (Fig. 2(c-d)). 
With an electric field, the dips move in the same direction 
in energy both for the nitrogen and for the boron impurity.
The shapes of the dips are distorted when the dips approach 
the Fermi level or subband edges.
The maximum resistances are about twice of the zero-field resistance 
because, at the corresponding electric fields, half of the conducting 
electrons at ${E}_F$ are backscattered by the impurities. 
Since the resonant backscatterings are shifted energetically
by the field-induced potential change at a defect site, 
the effect of the electric field scales as $|E_\text{ext}|\cos\theta$
approximately when the electric field ($E_\text{ext}$) has an angle
$\theta$ with the normal direction of the tubular surface 
at the defect site. So, the effect is maximized if $\theta=0$ as shown in Fig. 1(a). 
We have also examined, by first-principles calculations, 
electrical properties of single-impurity doped semi-metallic ($3n$, 0) carbon nanotubes 
which have very small energy gaps due to the curvature effect~\cite{ouyang}. 
Our calculation results show that, with transverse electric fields, the resistance of a doped 
(18,0) SWNT, which is measured at either the conduction or the valence band edge, 
has almost the same tunable behavior (not shown here).

By adding up the effects of the nitrogen and the boron impurity in a single nanotube,
we can produce much larger variation of the resistance so that 
with nitrogen and boron co-doping, a metallic carbon nanotube shows 
an ``on-to-off" switching behavior as a function of the applied electric field (Fig. 3(b)). 
When two carbon atoms on opposite sides of a (10, 10) nanotube are replaced 
by a nitrogen and a boron atom (Fig. 3(a)), the resistance of the nanotube is slightly (3\%) 
increased when no transverse electric field is present (Fig. 3(b)). 
It remains in the low-resistive ``on-state" (6.7 k$\Omega$). 
When an electric field of 0.4V/\AA~pointing from the nitrogen-doped 
side to boron-doped side (Fig. 3(a)) is applied, the resistance of 
the nanotube increases to 221 k$\Omega$, switching to a high-resistive ``off-state" (Fig. 3(b)). 
With further increase in the strength of the electric field, 
the resistance decreases back to 20 k$\Omega$. 
When the direction of the electric field is reversed, 
the doped nanotube remains in the low-resistive ``on-state" (less than 10 k$\Omega$)
at all values of field strength.

\begin{figure}[t]
\centering
\includegraphics[width=8.5cm]{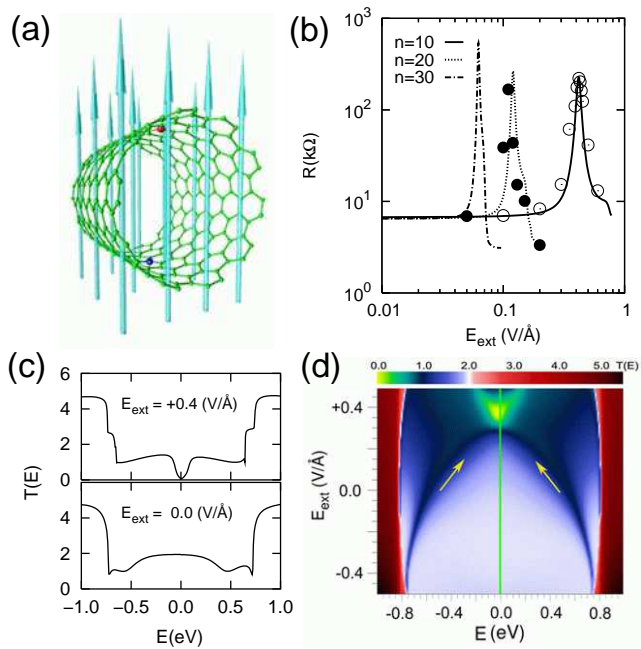}
\caption{(color online) Electrical resistance of nitrogen and boron co-doped 
SWNTs in transverse electric field $E_\text{ext}$. 
(a) A schematic diagram of a nitrogen (blue dot) and 
boron (red dot) co-doped SWNT in a transverse electric field. 
The direction of the arrows corresponds to $E_\text{ext}> 0$. 
(b) Calculated two-probe electrical resistances ($R$) 
of nitrogen and boron co-doped ($n$, $n$) SWNTs ($n$=10, 20, 30). 
Empty and filled circles are results from {\it ab-initio} 
calculations for (10,10) and (20,20) SWNTs, respectively, 
and lines are results from tight-binding approximations.  
(c) Transmission spectra $T(E)$ of the nitrogen-and-boron co-doped 
(10,10) SWNT in electric fields of 0.0 and 0.4 V/\AA.
(d) Color-scale plot of transmission spectra  $T(E)$  as a function of 
energy  $E$  and electric field  $E_\text{ext}$  
for the nitrogen-and-boron co-doped (10,10) SWNT. 
The color-scale bar is placed on top of the plot.}
\end{figure}

The high-resistive ``off-state" of the N and B co-doped metallic 
nanotubes originates from the impurity-induced resonant backscatterings 
which reflect most of the conducting electrons at ${E}_F$. 
With no electric field, resonant scatterings due to the N and 
the B impurity occur above and below ${E}_F$, respectively, 
reflecting half of the conducting electrons at their respective resonant energies (Fig. 3(c)). 
With nonzero $E_\text{ext}$,
the field-induced potential shifts are of the opposite sign near the two impurities 
since the two impurities are placed at opposite sides of the tube.
This makes resonant backscatterings by the two impurities 
move in opposite directions in energy (toward ${E}_F$) and, with a specific 
$E_\text{ext}$, they meet at ${E}_F$, causing almost complete 
reflection of the conducting electrons (Fig. 3(c)) 
and thereby the large increase of the resistance. 
In Fig. 3(d), it is shown that 
with electric fields, resonant backscattering dips (dark blue) 
associated with the boron and the nitrogen impurity move 
in opposite directions (denoted by yellow arrows) in energy 
because they are located on opposite sides of the nanotube. 
With $E_\text{ext}\simeq 0.4$V/\AA~in Fig. 3(d), 
the two dips merge, making transmission almost zero (yellow region) 
at $E_F$ (green vertical line).

The relative position of the boron and nitrogen impurities on the tubular surface affects
the magnitude of the maximum resistance at the ``off-state".
Let $z$ be the distance of the two impurities along the tubular axis,
and $\phi$ be their polar-angle difference.
When $\phi = 180 ^\circ$ and $z = (3n \pm 1)\times L$
for integral $n$ (as shown in Fig. 3(a)), 
the maximum resistance at the ``off-state" is about 221 k$\Omega$. 
Here, $L$ is the unit-cell length of the armchair-type nanotube. 
If $\phi = 0 ^\circ$ and $z = 3n\times L$, only one conduction 
channel is blocked at the maximal resistance condition, 
resulting in the maximal resistance of about 13 k$\Omega$.
These features are coherent effects of the two impurity scatterings, which occurs
if the two impurities are closer than the phase coherence length of the electrons.

In a larger-diameter SWNT, the electrical switching is possible 
with a weaker electric field. 
First-principles calculations are preformed for different diameter 
defective tubes up to the (20, 20) carbon nanotubes, and 
we adopt tight-binding calculations to even larger-diameter 
nanotubes such as (30, 30) and (40, 40) tubes (Fig. 3(b)).
The tight-binding approximation here adopts the standard one $\pi$-electron model 
with the nearest-neighbor hopping interaction ~\cite{fujita}.
A nitrogen (boron) impurity is represented by an attractive 
(repulsive) Gaussian onsite potential and the applied electric field is expressed 
with the onsite-energy variation screened with a dielectric constant.
Parameters for impurity potentials and the dielectric constant are fit to reproduce 
our first-principles calculations of the doped (10,10) nanotubes.
The required field strength to maximum resistance is found to be 
approximately inversely proportional to the square of the diameter 
of the nanotube. 
This scaling is due to the product of two effects: the field-induced 
potential drop across the tube diameter increases linearly and 
the subband energy interval decreases linearly with the increase 
of the diameter of the nanotubes~\cite{green}. 
For example, the ``off-state" for a nitrogen-boron co-doped (30, 30) 
nanotube can be reached by a field of 0.06 V/\AA~(Fig. 3(b)). 
Thus, the switching behavior will be more easily achieved 
in experiments with larger-diameter nanotubes.

\begin{figure}[t]
\centering
\includegraphics[width=8.5cm]{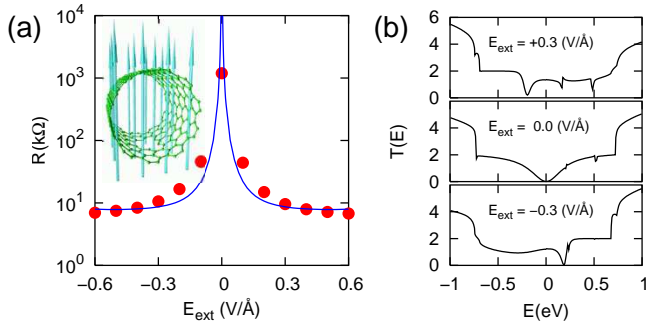}
\caption{(a) Calculated two-probe resistance of a 
(10,10) SWNT with a vacancy in transverse electric fields ($E_\text{ext}$). 
Filled circles result from {\em ab-initio} calculations while the 
solid line from a tight-binding approximation. 
The direction of the electric field in the inset represents 
$E_\text{ext}> 0$. 
(b) Transmission spectra ($T(E)$) of the (10, 10) SWNT 
with the vacancy as a function of energy ($E$). 
From the top to the bottom panel, the applied  
fields are $+$0.3, 0.0, and $-$0.3 V/\AA, respectively.}
\end{figure}

An ``off-to-on" switching behavior occurs when considering 
a carbon nanotube with a vacancy.
A recent observation on graphene layers and SWNTs~\cite{hashimoto} 
by transmission electron microscopy 
reveals stable intrinsic vacancy defects under strong electric field.
When four adjacent carbon atoms are removed in a (10, 10) SWNT 
(shown in the inset of Fig. 4(a)), for example, 
the nanotube is in a high-resistive ``off-state" with no electric field. 
Resonant backscatterings are at ${E}_F$ and 
the resistance is 1.2 M$\Omega$ (Fig. 4(a)). 
As a transverse electric field is applied, regardless of the sign of the field,
the resonant backscatterings move away from 
${E}_F$ 
(Fig. 4(b)) 
and the nanotube switches to a low-resistive ``on-state" 
having the resistance less than 20 k$\Omega$ for 
$|E_\text{ext}|\geq 0.12$V/\AA~
(less than 100 k$\Omega$ for $|E_\text{ext}|\geq 0.04$V/\AA)~\cite{vacant}.

In summary, we have shown theoretically that the resistance of SWNTs is tunable 
up to three orders of magnitude by both defects and transverse electric fields, 
even if the SWNTs are metallic. 
Such a result may open a new way to make 
electronic devices by exploiting the variation of the resistance 
in metallic and semiconducting nanotubes simultaneously 
as a function of the electric field.
The possible randomness of size of defects and those distributions
and the stability of defects under the electric fields will require 
future works~\cite{prel}.
We anticipate that this interesting property of the defective metallic SWNTs
can be tested by suitable experimental setup, e.g., isolated nanotube
in a nanoelectromechanical system~\cite{zettl} or 
in a split-gate configuration~\cite{sblee}.

Y.-W.S., M.L.C. and S.G.L. were supported by NSF Grant No. DMR04-39768 and 
by the Director, Office of Science, Office of Basic Energy under Contract
No. DE-AC03-76SF00098. J.I. was supported by the SRC program (Center
for Nanotubes and Nanostructured Composites) MOST/KOSEF. 
Computational resources have been provided by the
KISTI, NSF at the NPACI and DOE at the NERSC.


\begin{thebibliography}{99}
\bibitem{mceuen}P. L. McEuen, M. S. Fuhrer, H. Park, 
        IEEE transactions on nanotechnology {\bf 1}(1), 78 (2002).
\bibitem{tans}S. J. Tans, A. R. M. Verschueren, C. Dekker, Nature {\bf 393}, 49 (1998).
\bibitem{soh}H. T. Soh {\it et al}, Appl. Phys. Lett. {\bf 75}, 627 (1999). 
\bibitem{martel}R. Martel {\it et al}, Phys. Rev. Lett. {\bf 87}, 256805 (2001).
\bibitem{javey}A. Javey {\it et al}, Nature {\bf 424}, 654 (2003).
\bibitem{bockrath}P. L. McEuen {\it et al}, Phys. Rev. Lett. {\bf 83}, 5098 (1999).
\bibitem{nygard}J. Nyg{\aa}rd {\it et al}, Appl. Phys. A {\bf 69}, 297 (1999).
\bibitem{okeeffe}J. O'Keeffe, C. Wei, K. Cho, Appl. Phys. Lett. {\bf 80}, 676 (2002).
\bibitem{li}Y. Li, S. V. Rotkin, U. Ravaioli, Nano Lett. {\bf 3}, 183 (2003).
\bibitem{novikov}D. S. Novikov, L. S. Levitov, cond-mat/0204499.
\bibitem{yhkim} Y.-H. Kim, K. J. Chang, Phys. Rev. B {\bf 64}, 153404 (2001).
\bibitem{barraza}S. Barraza-Lopez {\it et al}, Europhys. Lett. {\bf 69}(6), 1003 (2005).
\bibitem{choi2}H. J. Choi {\it et al}, Phys. Rev. Lett. {\bf 84}, 2917 (2000).
\bibitem{kostyrko}T. Kostyrko, M. Bartkowiak, G. D. Mahan, Phys. Rev. B {\bf 59}, 3241 (1999).
\bibitem{ando}T. Ando, J. Phys. Soc. Jpn. {\bf 74}, 777 (2005). 
\bibitem{zhao}J. Zhao, R. Xie, J. Nanosci. Nanotech. {\bf 3}, 459 (2003).
\bibitem{saito}R. Saito, G. Dresselhaus, M. S. Dresselhaus, 
        {\it Physical Properties of Carbon Nanotubes} 
        (Imperial College Press, London, 1998).
\bibitem{frank}S. Frank {\it et al}, Science {\bf 280}, 1744 (1998).
\bibitem{landauer}R. Landauer, Philos. Mag. {\bf 21}, 863 (1970).
\bibitem{fisher}D. S. Fisher, P. A. Lee, Phys. Rev. B {\bf 23}, R6851 (1981).
\bibitem{benedict}L. X. Benedict, S. G. Louie, M. L. Cohen, 
         Phys. Rev. B {\bf 52}, 8541 (1995).
\bibitem{soler}J. M. Soler {\it et al}, J. Phys.: Cond. Mater. {\bf 14}, 2745 (2002).
\bibitem{troullier}N. Troullier, J. L. Martins, Phys. Rev. B {\bf 43}, 1993 (1991).
\bibitem{kleinman}L. Kleinman, D. M. Bylander, Phys. Rev. Lett. {\bf 48}, 1425 (1982).
\bibitem{choi1}H. J. Choi, M. L. Cohen, S. G. Louie, {\it to be published}.
\bibitem{yyoon}Y.-G. Yoon {\it et al}, Phys. Rev. Lett. {\bf 86}, 688 (2001).
\bibitem{datta}S. Datta, {\it Electronic Transport in Mesoscopic Systems} 
               (Cambridge Univ. Press, 1995).
\bibitem{ouyang}M. Ouyang {\it et al}, Science {\bf 292}, 702 (2001). 
\bibitem{green}The approximate scaling behavior
can be shown from the Lippmann-Schwinger formalism.
The scattered wave $|\psi\rangle$ by the impurity is 
$|\psi\rangle=|\varphi\rangle+\hat{G}\hat{V}_N|\psi\rangle$
with the incoming state $|\varphi\rangle$, 
the Green function $\hat{G}$,
and the nitrogen impurity potential $\hat{V}_N$. 
The applied $E_\text{ext}$ changes 
$\hat{G}$ to the first order by
$\delta\hat{G}=\sum_n \frac{|\varphi_n\rangle\langle\delta\varphi_n |
+|\delta\varphi_n\rangle\langle\varphi_n |}{E-E_n }$, which
scales as $|{\bf E}_\text{ext}|r^2$ because 
$E-E_n \propto 1/r$  and
$|\delta\varphi_n\rangle\langle\varphi_n |\simeq \sum_{m=n\pm 1}
\frac{\langle\varphi_n |{\bf E}_\text{ext}\cdot{\bf z}|\varphi_m\rangle}{E_n-E_m}
|\varphi_m\rangle\langle\varphi_n |
\propto \frac{|{\bf E}_\text{ext}|r(1/\sqrt{r})^2}{1/r} =  |{\bf E}_\text{ext}|r$.
Here, $r$ is the radius of a SWNT, and we used
$E_n-E_m \propto 1/r$~\cite{saito} and the selection rule ($m-n=\pm 1$) 
of the subband mixing~\cite{yhkim}.
\bibitem{fujita}R. Saito {\it et al}, Phys. Rev. B {\bf 46}, 1804 (1992).
\bibitem{hashimoto}A. Hashimoto {\it et al}, Nature {\bf 430}, 870 (2004).
\bibitem{vacant}
Vacancy defects with hydrogen passivations to $\sigma$ bonds 
are found to provide a single or multiple 
quasi-bound states from $\pi$ orbitals around ${E}_F$ depending on 
the number of removed atoms. 
The detailed calculations on various vacancy defects will be published elsewhere.
\bibitem{prel}
Preliminary first-principles calculations were performed to show similar 
switching behaviors in (10,10) SWNT doped with multiple(two and three) nitrogens, 
two boron and two nitrogen, and three boron and one nitrogen substitutional 
impurities in various configurations.
\bibitem{zettl}A. M. Fennimore {\it et al}, Nature {\bf 424}, 408 (2003).
\bibitem{sblee}S.-B. Lee {\it et al}, J. Nanosci. Nanotech. {\bf 3}(4), 325 (2003).
\end{thebibliography}
\end{document}